\documentclass{LMCS}

\def\doi{8 (3:09) 2012}
\lmcsheading%
{\doi}
{1--8}
{}
{}
{Feb.~22, 2012}
{Aug.~13, 2012}
{}

\usepackage{enumerate,hyperref}

\newcommand\np{{\mbox{NP}}}
\newcommand\pp{{\mbox{P}}}
\newcommand\npco{{\np \cap co\np}}
\newcommand\nn{{\{0,1\}^n}}
\newcommand\mm{{\{0,1\}^m}}

\newcommand\el{{\{0,1\}^{\ell}}}

\newcommand\lz{[\mkern-2.5mu[}
\newcommand\rz{]\mkern-2.5mu]}
\newcommand\lb{\langle\mkern-2.5mu\langle}
\newcommand\rb{\rangle\mkern-2.5mu\rangle}

\begin{document}

\title[Pseudo-finite hard instances]%
{Pseudo-finite hard instances for a student-teacher
game with a Nisan-Wigderson generator}

\author[J.~Kraj\'{\i}\v{c}ek]{Jan Kraj\'{\i}\v{c}ek}

\address{Faculty of Mathematics and Physics\\
Charles University in Prague\newline
Sokolovsk\' a 83, Prague 8, CZ - 186 75\\
The Czech Republic}

\email{krajicek@karlin.mff.cuni.cz}

\thanks{Supported
in part by grant IAA100190902.
Also partially affiliated with the Institute of Mathematics of the Academy of Sciences.}


\begin{abstract}
  \noindent 
For an $\np \cap co\np$ function $g$ of the Nisan-Wigderson type and a string $b$ outside its range we consider
a two player game on a common input $a$ to the function.
One player, a computationally limited Student, tries to 
find a bit of $g(a)$ that differs from the corresponding bit of $b$. 
He can query a computationally unlimited Teacher
for the witnesses of the values of constantly many
bits of $g(a)$. The Student computes the
queries from $a$ and from
Teacher's answers to his previous queries.  

It was proved in [Kra11b] that if $g$ is based 
on a hard bit of a one-way permutation
then no Student computed by a polynomial size circuit can succeed on all $a$. 
In this paper we give a lower bound on the number
of inputs $a$ any such Student must fail on.
Using that we show that there is a pseudo-finite 
set of hard instances on which all uniform students
must fail. The hard-core set is defined 
in a non-standard model of true arithmetic and has applications
in a forcing construction from [Kra11a].  
\end{abstract}

\maketitle

\section*{Introduction}

Consider a function $g : \nn \rightarrow \mm$ defined as a Nisan-Wigderson
generator based on some Boolean function $f$, cf.\cite{NW}. That is, there is a set system
$\{J_i\subseteq [n]\}_{i\in [m]}$ such that

\begin{enumerate}[$\bullet$]

\item $|J_i| = \ell$, for all $i$

\item $|J_i \cap J_j| \le d$, for all $i \neq j$

\end{enumerate}
and the $i$-th bit of $g(x)$ equals to $f(x(J_i))$ where $f : \el \rightarrow \{0,1\}$
and $x(J_i)$ is the $\ell$-bit string
$x_{j_1}\dots x_{j_{\ell}}$ if
\[
J_i\ =\ \{j_1< \dots < j_{\ell}\}\ .
\]
We are interested in the case when $f$ is a hard bit $B(v)$ of a
polynomial-time computable one-way permutation $h : \el \rightarrow \el$:
\[
f(u)\ :=\ B(h^{(-1)}(u))\ .
\]
If $m > n$ there are strings $b \in \mm \setminus Rng(g)$
and with any such string $b$ we associate the following game
${mathcal G}_b$. The game is played by two players, a Student and a Teacher,
both knowing $b$.
They receive a common input: any $a \in \nn$. The Student tries to find
$i \in [m]$ such that $g(a)_i \neq b_i$, certifying thus that
$b$ is outside of the range of $g$. The Student will be computationally
limited and, in particular, while he will be able to compute permutation $h$ 
he will not be able to compute function $f$
and to find a suitable bit $i$ himself. Instead he will compute some
candidate solution $i_1$ and hand it to the Teacher. She has
an unlimited computational power and will reply to the Student with
the unique $v^1 \in \el$ such that $h(v^1) = a(J_{i_1})$.

If $B(v^1) \neq b_{i_1}$ the game stops with the Student solving the task.
Otherwise he computes his next candidate solution $i_2 \in [m]$ and
hands it to the Teacher, gets back $v^2$ such that $h(v^2) = a(J_{i_2})$,
etc.

In general the Student will be allowed to present up to $c$ 
candidate solutions, $c$ some parameter. If he does not find
a solution, we say he failed. The Student can be modelled
by $c$ functions
\[
S_1(x), S_2(x, y^1), \dots, S_c(x, y^1, \dots, y^{c-1})
\]
computing his candidate solutions:
\[
i_k \ :=\ S_k(a, v^1, \dots, v^{k-1}), \ k \le c
\]
where $v^j$'s are the Teacher's replies.

We are interested in the case when the Student is a small circuit, i.e.
the functions $S_k$ are computed by circuits of a small total size.
The Teacher's moves are uniquely determined. This Student-Teacher
way of computing a function grew out of a use of Herbrand's theorem 
in bounded arithmetic; the Student-Teacher formalism was introduced
in \cite{KPS} (see there also
for an overview in a complexity-theoretic language).

\bigskip

Assuming that $h$ is indeed a one-way permutation it was proved
in \cite{Kra-nwg,k2} that for any fixed $c \geq 1$ no 
$\pp/poly$ Student can solve the task on all inputs $a \in \nn$,
for $n >> 0$. In this paper we are interested in the question
whether there exists a set of hard instances $H \subseteq \nn$
such that every $\pp/poly$ Student fails on most 
$a \in H$.

This question is motivated by a research in proof complexity;
in particular by a conjecture\footnote{Statement (S)
in \cite[Sec.31.4]{k2} or in \cite{Kra-nwg}, modifying  Razborov's
Conjecture 2 from \cite{Raz03} and 
related also to Rudich's demi-bit conjecture
from \cite{Rud}.}
that functions like $g$ are hard
proof complexity generators and by a model-theoretic approach to it
based on forcing. This proof complexity motivation is 
discussed in detail in \cite[Part VIII]{k2} and in \cite{Kra-nwg}
and we shall not review it here.

\bigskip

In this paper we show that we can combine the lower bound argument
from \cite[Chpt.31]{k2} and \cite{Kra-nwg} with the 
model-theoretic set-up of forcing with random variables
to give a partial affirmative answer to the question about the existence
of sets $H$. The qualification "partial" means that we
construct such a set of hard examples as a pseudo-finite set
of strings of a non-standard length $n$
in a model of true arithmetic and that the instances 
from it are hard for uniform students, i.e. the functions
$
S_1(x), S_2(x, y^1), \dots, S_c(x, y^1, \dots, y^{c-1})
$
defining their moves are computed by uniform algorithms. In fact,
we can allow the uniform students to use 
common advice strings of all lengths. This partial solution is perfectly adequate
for the purpose of the construction in 
\cite[Chpt.31]{k2}\footnote{The original construction was based
on an unsuitable sample space as pointed out by S.~Buss.
This is explained in the last section.}  
but it would be desirable to have such hard-core sets for non-uniform
students as well. This is because the forcing models constructed
in that case have some nice properties (in particular, witnessing
of quantifiers \cite[Chpt.3]{k2} or saturation properties
\cite{Kra-satur}) that may be useful for further research.

\bigskip

The paper is self-contained in the sense that the reader will be able to 
fully understand the problem and its solution. However, to appreciate
its relevance to the forcing construction and subsequently for
the proof complexity
conjecture alluded to above one needs to consult \cite[Part VIII]{k2}
or \cite{Kra-nwg} (better both).

The paper is organized as follows. In Section \ref{one} we give details
of the problem and use the argument from \cite{Kra-nwg,k2} 
to derive a lower bound on the number of inputs every $\pp/poly$ student
must fail on. In Section \ref{two} some model theory
and forcing with random variables is briefly reviewed and the
pseudo-finite hard-core set is constructed in Section \ref{three}.

Relevant background can be found in \cite{kniha,k2,Kra-nwg}. 
In particular, readers not familiar with non-standard models can find 
a self-contained introduction to the topic in
the appendix in \cite{k2}.

\section{Hardness of the game}
\label{one}

We shall fix the following parameters in the definition of
function $g$:
\[
m:=n+1,\ \ \ \ \ell := n^{1/3},\ \ \ \ \ d := \log(m)\ .
\]
In applications of Nisan-Wigderson generators $m$ is usually
exponentially large but for the purpose of proof complexity
the best choice is to have $m$ as small as possible.
By \cite{NW} there is a set system $\{J_i\}_i$
with the required properties, and we fix any one of them. 

Let $h$ be a polynomial time permutation (we are interested in its
restriction to $\el$) that is one-way and let $B(v)$ be
its hard bit. In particular, by this hardness assumption we mean that
\begin{enumerate}[$\bullet$]

\item
For any fixed
$k \geq 1$ no $\pp/poly$ algorithm can compute from $u \in \el$
the value $B(h^{(-1)}(u))$ with the advantage better than
$\ell^{-k}$ over $1/2$.

\end{enumerate}
\bigskip

Let us fix a string $b \in \mm \setminus Rng(g)$.
In \cite{Kra-nwg} (a similar argument is used in \cite[Sec.31.2]{k2}
for a different purpose) it was proved that no $\pp/poly$
Student can succeed
in the game ${mathcal G}_b$ on all inputs, assuming that $h$ is indeed
one-way. We are now going to sketch the argument in order
to extend it a bit and to
deduce a lower bound on the number of inputs on which
a Student must fail. Any details of the original
argument missing here can be found in the proof
of \cite[Thm.3.2]{Kra-nwg}.

\bigskip

Consider
a Student that attempts to succeed in the game ${mathcal G}_b$
on as many inputs $a$ as possible and assume he can ask at most
$c$ queries to the Teacher. The lower bound will depend on $c$.

Denote by $W \subseteq \nn$
the set of all $a \in \nn$ on which the Student succeeds. 
For $a \in \nn$ let us denote by
$v^i(a)$ the preimage of $a(J_i)$ in $h$, for $i \in [m]$.

If the Student succeeded on $a$ and $i$ was its last query to the Teacher
it means that
\[
f(a(J_i)) \neq b_i\ .
\]
Intuitively this gives us some information about the function $f$
as we can deduce its value on the string $a(J_i)$ while receiving during the 
computation from the Teacher only strings that have 
little to do
with the string $h^{(-1)}(a(J_i))$ we would need in order to
compute $f$ ourselves. The formal argument makes this intuition
precise. 

\bigskip

Assume that the Student asked $k$ queries:
the candidate solutions the Student produced were 
$i_1, \dots, i_k$ and the last one $i_k$ was correct. Call
the $k$-tuple $(i_1, \dots, i_k)$ the trace of the computation on $a$. 
In particular, $k \le c$.
As the witnesses are unique the trace determines 
also the Teacher's replies. A simple counting argument yields
(cf. the proof of \cite[Thm.3.2]{Kra-nwg})

\medskip
\noindent
{\bf Claim 1:} {\em There is a $k$-tuple $(i_1, \dots, i_k)
\in [m]^k$
for some $k \le c$ that is the trace of computations on at least a
fraction of $\frac{2}{(3m)^k}$ of all inputs from $W$.}

\medskip

Fixing a trace ${\overline i} = (i_1, \dots, i_k)$ 
satisfying the claim, define
for any $u \in\{0,1\}^{\ell}$ and $v \in \{0,1\}^{n - \ell}$
the string $a(u,v) \in \nn$ as follows: put bits of $u$ into the positions
$J_{i_k}$ and then fill the remaining $n - \ell$
positions by bits of $v$. The following claim follows from the proof
of Claim 1 by averaging.

\medskip
\noindent
{\bf Claim 2:} {\em There is $e \in \{0,1\}^{n - \ell}$
s.t. at least a fraction of $\frac{1}{(3m)^k}$
more $u \in \el$ yield sample $a(u,e) \in W$
whose trace is exactly $\overline i$ than those $u$ which
yield $a(u,e) \in W$ whose trace properly contains $\overline i$.}

\medskip

Fix one such an $(n-\ell)$-tuple $e$. Call any $u \in \el$ good if 
$a(u,e) \in W$.

\bigskip

The property that two distinct sets from the set system $\{J_i\}_i$
intersect in at most $\log(m)$ positions implies that there are,
for any row $i \neq i_k$, at most $m$ assignments $w$ to bits in
$J_i$ not set by $e$. Any such $w$ determines, together with $e$, an
assignment to variables in $J_i$ and hence a string in $\{0,1\}^{\ell}$;
denote it $z_w$. Let $Y_i$ be the set of all preimages of all 
$z_w$ in the permutation $h$. The 
total bit size of all $Y_i$ together is $m^{O(1)}$.

\medskip

This situation allows us to define an algorithm $C$ that uses as
advice the set system $\{J_i\}_i$, the string $b$,
the trace ${\overline i}$, the
partial assignment $e$, and all $m-1$ sets $Y_i$.
The total size of the advice is again bounded above by
$m^{O(1)}$.

The algorithm $C$ attempts to compute the function $f$ on inputs $u \in \el$.
Let $U$ be those inputs $u \in \el$ for which the trace of $a(u,e))$ either 
equals to ${\overline i}$ or starts with ${\overline i}$,
and let $b_0$ be the majority value of $f$ on the complement of $U$.

On input $u \in \{0,1\}^{\ell}$ $C$ simulates the Student's computation 
on the string $a : = a(u,e) \in \nn$.
If any of the candidate solutions produced in the $j$-th query, 
$j = 1, \dots, k$, to the Teacher differs
from the $j$-th entry in the trace $\overline i$ 
$C$ halts and outputs $b_0$. 

Otherwise, if the trace of the computation follows $\overline i$,
$C$ uses sets $Y_i$ to simulate Teacher's replies (these are unique
and can be tested as correct).
If the computation evolved according to the trace ${\overline i}$
and reached the $k$-th step $C$ outputs $1 -  b_{i_k}$. 

The algorithm $C$ outputs the bit $b_0$ in all cases except 
when the computation follows the trace $\overline i$ and
reaches the $k$-th step.
If the computation of the Student were to actually stop 
at that point then the value 
$1 -  b_{i_k}$ is indeed equal to $f(u)$. 

Otherwise, if the computation
were to continue, we do not have a way to deduce the 
true value of $f(u)$. 
The influence of this case can be, however, bounded.
By the choice of $e$ in Claim 2 the former
case happens for at least a fraction of
$\frac{1}{(3m)^k}$ more of all good inputs $u \in \{0,1\}^{\ell}$
than the latter one.
If $W = \nn$ (as in \cite{Kra-nwg}) we would be done:
All $u$ are good and as
$b_0$ is the correct value of $f$ for at least half of
$u \notin U$, the algorithm $C$ would compute $f$
with an advantage over $1/2$ at least $\frac{1}{(3m)^k}$.

\bigskip

If $W$ is a proper subset of $\nn$ this argument fails as we
have no control over the number of bad $u$ (i.e. for which 
$a(u,e) \notin W$) but the trace of $a(u,e)$ contains $\overline i$, 
i.e. of the size of
the set $U \setminus W$.
However, if we knew that the size of the complement of $W$
in $\nn$ is at most, say:
\[
s_c\ :=\ \frac{1}{2}\cdot 2^{n^{1/3}}\cdot \frac{1}{(3m)^c}
\]
then the above argument works: $s_c$ bounds, in particular, 
the number of bad $u$
and the algorithm $C$ gets the advantage
at least
\[
\frac{1}{(3m)^k} -
\frac{1}{2}\cdot  \frac{1}{(3m)^c}\ 
\geq\ 
\frac{1}{2}\cdot \frac{1}{(3m)^k}\ \geq\
\frac{1}{2}\cdot \frac{1}{(3m)^c}\ . 
\]
The algorithm $C$ needs the same
time as the Student except when it simulates a reply of the Teacher
and looks for an appropriate witness in one of the sets $Y_i$. 
This is done at most $(c-1)$-times and takes $m^{O(1)}$ time per
one witness-search. 
Hence if the Student is $P/poly$ the total time 
$C$ uses is $c\cdot m^{O(1)}$.

\bigskip

We conclude that assuming that a $\pp/poly$ Student fails on less
than $s_c$ inputs 
contradicts the hypothesis that $h$ is a one-way permutation.
Hence the following statement was established.

\begin{lem}\label{2.1}
Assume that the parameters $n, m , \ell$, the set system $\{J_i\}_i$
and the string $b$ satisfy the conditions imposed on them
earlier. Assume also that $h$ is a polynomial-time one-way permutation, that
$B(v)$ is its hard bit and that $f(u) = B(h^{(-1)}(u))$.

Let $c \geq 1$ be arbitrary. Then for any fixed $k \geq 1$,
for any $n$ large enough any Student 
asking at most $c$ queries to the Teacher and computed
by circuits of the total size $m^k$ must fail on at least
\[
s_c\ :=\ \frac{1}{2}\cdot 2^{n^{1/3}}\cdot \frac{1}{(3m)^c}
\]
inputs $a \in \nn$.
\end{lem}

Let us remark that if the hardness of the permutation $h$
were exponential, in the sense that for some $\epsilon > 0$
a circuit needs to have the size at least $2^{\ell^{\epsilon}}$
in order to compute the hard bit with an advantage at least
$2^{- \ell^{\epsilon}}$, then we could allow also exponentially
large Students posing up to $n^{\delta}$ queries for some small
$\delta > 0$ (depending on $\epsilon$)
and still get a meaningful lower bound. 
Also, the whole situation can be specialized to
various circuit subclasses of $\pp/poly$ such as $AC^0$ or $NC^1$,
as discussed in \cite{Kra-nwg}.

\section{Some model theory}
\label{two}

In this section we briefly recall the set-up of forcing with random
variables, enough to formulate and prove our result in the next section.
However, we shall not go into the details specific for the construction
in \cite[Chpt.31]{k2} motivating this paper.

Forcing of random variables is a method how to construct models
of arithmetic. A special emphasis is given to bounded arithmetic because
of its relation to proof complexity but the method is not limited
to this theory. The models are formed from random variables
on a pseudo-finite sample space and are Boolean-valued.

Let $mathcal M$ be a non-standard $\aleph_1$-saturated
model of true arithmetic in some language $L$
containing the language of Peano arithmetic
and having a canonical interpretation in the standard model. 
Let $\Omega \in {mathcal M}$
be an infinite set; as it is an element of the model it is finite
from the point of view of $mathcal M$. 
Let $F \subseteq {mathcal M}$ be any - not necessarily definable - family 
of functions $\alpha : \Omega \rightarrow {mathcal M}$.

The family $F$ will be the universe of a Boolean-valued $L$-structure
$K(F)$. The symbols of $L$ are interpreted by composition
with functions from $F$.
For example, for a $k$-ary function symbol $f$ and any
$\alpha_1, \dots , \alpha_k \in F$ define
the function $f(\alpha_1, \dots , \alpha_k)$ by
\[
f(\alpha_1, \dots , \alpha_k)(\omega)\ :=\ 
f(\alpha_1(\omega), \dots , \alpha_k(\omega)), \ \mbox{ for } 
\omega \in \Omega\ .
\]
We need to assume that this function is also in $F$, i.e.
that $F$ is $L$-closed
in the terminology of \cite{k2}.

Every atomic $L(F)$-sentence $A$ is naturally assigned a subset
$\lb A \rb \subseteq \Omega$ consisting of those samples 
$\omega \in \Omega$ for which $A$ is true in ${mathcal M}$.

Combining the idea of Loeb's measure with some measure
theory
it was shown in \cite[Sec.1.2]{k2} that if we factor the Boolean algebra
of $mathcal M$-definable subsets of $\Omega$ by the ideal of sets
of an infinitesimal counting measure we get a complete Boolean algebra
$mathcal B$.

The image of $\lb A \rb$ in $mathcal B$ in this quotient is denoted
$\lz A \rz$. Following Boole \cite{Boo} and Rasiowa-Sikorski \cite{RS1}
this determines the truth value $\lz A \rz \in {mathcal B}$
for any $L(F)$-sentence $A$: $\lz \dots \rz$ commutes with
Boolean connectives and
\[
\lz \exists x A(x)\rz\ :=\ \bigvee_{\alpha\in F} \lz A(\alpha)\rz
\]
and analogously for the universal quantifier.

There are various generalizations of this basic set-up
and, in particular, the random variables from the family $F$
can be only partially defined on the sample space $\Omega$,
as long as their regions of undefinability have infinitesimal counting
measures.

In the particular construction in \cite[Chpt.31]{k2}
the sample space is simply the set $\nn$ for some non-standard
$n \in {mathcal M}$, and the family $F$ consisted of partial
random variables computed by students operating similarly as in the game 
${mathcal G}_b$. More precisely, any partial 
function $\alpha \in F$ is computed by 
a $\pp/poly$ Student who 
\begin{enumerate}[$\bullet$]

\item gets an input $\omega \in \nn$,

\item there is a standard parameter $c$ such that the
Student can ask the Teacher for the values
of $h^{(-1)}$ on $\omega(J_i)$ for up to $c$ values of $i \in [m]$
($c$ is common to all inputs $\omega$ but 
may differ for different $\alpha$),

\item if the Teacher's answer $v$ to any query about
$h^{(-1)}(\omega(J_i))$ does not satisfy $B(v) = b_i$ 
the computation is aborted and $\alpha$ is undefined,

\item if the computation is not aborted after any Teacher's answer
the Student outputs at the end an element of ${mathcal M}$ 
(necessarily of size polynomial in $n$).

\end{enumerate}
Unfortunately such a function $\alpha$ is typically undefined on a
(standard) positive fraction\footnote{Contrary to what was claimed
in \cite[L.31.2.1]{k2} - I am indebted to S.~Buss for pointing it out.
See 
{\tt http://www.karlin.mff.cuni.cz/$\tilde{\ }$krajicek/k2-upravy.html}
for an explanation and a correction.}
on $\nn$ and hence the pair $\nn, F$ does not conform to the set-up
of the construction.

What is needed is an infinite subset $H \subseteq \Omega$, definable
in ${mathcal M}$, such that every student-computed $\alpha$ 
is defined on all but an infinitesimal fraction of $H$.

In the next section we shall construct such $H$ for uniform students
that may all use some common advice string. This is perfectly sufficient for
the intended applications of the model.
But it would still be desirable to have such a hard-core set for
non-uniform students of some superpolynomial size.
The reason is that a family $F$ based on such students can be 
modified into a compact one (in the sense of \cite[Sec.3.4]{k2}) and the
compactness of $F$ implies that the resulting model $K(F)$ has some nice 
model-theoretic properties mentioned in the introduction. For example, 
non-uniform students of sub-exponential size (i.e.
the functions 
$
S_1(x), S_2(x, y^1), \dots, S_c(x, y^1, \dots, y^{c-1})
$
defining their moves are computed by circuits of total size
$2^{n^{o(1)}}$) define family $F$ that is already compact.

\section{A construction of a hard-core set}
\label{three}

We assume that a 
non-standard $n \in {mathcal M}$,
a set system $\{J_i\}_{i}$ with the required properties, 
a permutation $h$ and its hard bit $B(v)$, 
and some string $b \in \mm \setminus Rng(g)$ are fixed.

Let $w \in {mathcal M}$ be any string of size polynomial in $n$
and let $F^{unif}_{w} \subseteq {mathcal M}$ be
the family of partial random variables on $\nn$ 
defined as $F$ in Section \ref{two} but allowing all Students computing 
the random variables to use as an advice
only the triple $(\{J_i\}_{i},b,w)$. This is perfectly sufficient for any
application\footnote{Note that \cite{Kra-nwg} already provided
an alternative construction with the same consequences as
those described in \cite{k2}.} of the eventual model 
in Sections 31.3. and 31.4 of
\cite{k2} ($w$ can contain e.g. a proof of a propositional formula or a witness
of the membership of $b$ in an $\np$ set, etc.)
and has the great advantage that the family $F_{w}^{unif}$ is now
countable.

\begin{thm}
There exists an infinite set $H \subseteq \nn$, 
$H \in {mathcal M}$, such that each $\alpha \in F^{unif}_{w}$ 
is defined on all samples from $H$.
\end{thm}

\proof

Enumerate as $\alpha_1, \alpha_2, \dots $ the set $F^{unif}_{w}$
in such a way that the Student defining $\alpha_k$
runs in time $\le m^k$ and asks at most $k$ queries, for all $k \geq 1$.

By the $\aleph_1$-saturation there exists a sequence
in ${mathcal M}$ of a non-standard length $t$ whose $k$-th element 
is $\alpha_k$, for all standard $k$ (see \cite[p.9]{k2}).
We shall denote it suggestively 
$(\alpha_i)_{i < t}$.

If we take $\alpha_1, \dots, \alpha_k$ we can compose the Students
defining them by first computing $\alpha_1$, if it is not aborted
then instead of outputting a value computing $\alpha_2$, etc. , and outputting 
(arbitrary) values only at the end, if the computation is not aborted earlier.
The resulting function is computed in time $O(k m^k)$ using at most
$k(k+1)/2 \le k^2$ queries. Hence by Lemma \ref{2.1} it is defined on
at least $s_{k^2}$ samples from $\nn$. This yields the following

\medskip

\noindent
{\bf Claim:} {For each standard $k \geq 1$ there exists definable
subset $H^k \subseteq \nn$ of size at least $s_{k^2}$
such that all $\alpha_1, \dots, \alpha_k$ are defined on all samples
from $H^k$.}

\medskip

\noindent
By the Overspill the statement of the Claim holds also for
the sequence $(\alpha_i)_{i \le r}$ for some non-standard
$r \le t$, and we can take $r$ small enough (but still non-standard)
such that $s_{r^2}$ is non-standard 
and hence the set $H := H^r$ satisfies the statement of the theorem.

\qed

Let us remark in conclusion that proofs of the Boolean case
hard-core set theorem of Impagliazzo \cite{Imp} do not seem to
work here. This is because $n$ small Students cannot be combined
into a one somewhat larger as this would blow-up
the number of queries posed to the Teacher.

\section*{Acknowledgment}
  \noindent  I am indebted to the two anonymous referees
for comments and suggestions.


\begin{thebibliography}{444444}

\bibitem[Boo47]{Boo}
G.~Boole.
\newblock {\em The mathematical analysis of logic}.
\newblock Barclay and Macmillan, Cambridge, 1847.

\bibitem[Imp95]{Imp}
R.~Impagliazzo.
\newblock Hard-Core Distributions for Somewhat Hard Problems.
\newblock
36th Annual Symp. on Foundations of Computer Science, 
Milwaukee, Wisconsin, 23-25 October 1995, IEEE Computer Society, 
538--545, 1995.

\bibitem[Kra95]{kniha}
J.~Kraj\'{\i}\v cek.
\newblock {\em Bounded arithmetic, propositional
logic, and complexity theory}.
\newblock  Encyclopedia of Mathematics
and Its Applications, Cambridge University Press,
Vol.60, 1995.

\bibitem[Kra11a]{k2}
J.~Kraj\'{\i}\v cek.
\newblock {\em Forcing with random variables 
and proof complexity}.
\newblock
London Mathematical Society
Lecture Notes Series, Cambridge University Press, Vol.382, 2011.

\bibitem[Kra11b]{Kra-nwg}
J.~Kraj\'{\i}\v{c}ek.
\newblock 
On the proof complexity of the Nisan-Wigderson generator
based on a hard $\npco$ function.
\newblock
{\em J. Mathematical Logic}, 11(1): 
11--27, 2011.

\bibitem[Kra??]{Kra-satur}
J.~Kraj\'{\i}\v{c}ek.
\newblock 
A saturation property of structures obtained
by forcing with a compact family of random variables.
\newblock
submitted preprint, 2012.


\bibitem[KPS90]{KPS}
J.~Kraj\'\i\v{c}ek, P.~Pudl\'ak, and
J.~Sgall.
\newblock
Interactive
Computations of Optimal Solutions.
\newblock
 in: B. Rovan (ed.): 
{\em Mathematical
Foundations of Computer Science} 
(B. Bystrica, August '90), Lecture
Notes in Computer Science, 
Springer-Verlag, 452:48--60, 1990.


\bibitem[NW94]{NW}
{N.~Nisan} and {A.~Wigderson}.
\newblock 
Hardness vs. randomness.
\newblock
 {\em J. Comput. System Sci.}, 
49:149--167, 1994. 

\bibitem[RS53]{RS1}
H.~Rasiowa and R.~Sikorski.
\newblock
Algebraic treatment of the notion of satisfiability.
\newblock
{\em Fundamenta Mathematicae}, 40:62--65, 1953..

\bibitem[Raz??]{Raz03}
A.~A.~Razborov.
\newblock
Pseudorandom generators hard for $k$-DNF resolution
and polynomial calculus resolution.
\newblock
unpublished preprint, 2003.

\bibitem[Rud97]{Rud}
S.~Rudich.
\newblock
 Super-bits, demi-bits, and
$\tilde{NP}/qpoly$-natural proofs.
\newblock
 in: {\em Proc.
of the 1st Int.Symp. on Randomization and Approximation 
Techniques in Computer Science}, LN in Comp.Sci., 
Springer-Verlag, 1269:85--93, 1997.

\end{thebibliography}
\end{document}